\documentclass[showpacs,preprintnumbers,amsmath,amssymb,superscriptaddress,nofootinbib,pra]{revtex4-1}

\usepackage{graphicx}
\usepackage{amsmath}
\usepackage{epsfig}
\usepackage{bm}
\usepackage{cases}
\usepackage{empheq}
\usepackage[usenames]{color}

\def\>{{\rangle}}
\def\<{{\langle}}

\begin{document}
   
\title{Dynamical Casimir effect in dissipative optomechanical cavity interacting with photonic crystal} 

\author{Satoshi \surname{Tanaka}}
\email{stanaka@p.s.osakafu-u.ac.jp}
\affiliation{Department of Physical Science, Osaka Prefecture 
University, Gakuen-cho 1-1, Sakai 599-8531, Japan}
\author{Kazuki \surname{Kanki}}
\affiliation{Department of Physical Science, Osaka Prefecture 
University, Gakuen-cho 1-1, Sakai 599-8531, Japan}

%
%
%
%
%
%
%


\begin{abstract}
We theoretically study the dynamical Casimir effect (DCE), i.e., parametric amplification of a quantum vacuum, in an optomechanical cavity interacting with a photonic crystal, which is considered to be an ideal system to study the microscopic dissipation effect on the DCE.
Starting from a total Hamiltonian including the photonic band system as well as the optomechanical cavity, we have derived an effective Floquet-Liouvillian by applying the Floquet method and Brillouin-Wigner-Feshbach projection method.
The microscopic dissipation effect is rigorously taken into account in terms of the energy-dependent self-energy.
The obtained effective Floquet-Liouvillian exhibits the two competing instabilities, i.e., parametric and resonance instabilities, which determine the stationary mode as a result of the balance between them in the dissipative DCE.
Solving the complex eigenvalue problem of the Floquet-Liouvillian, we have determined the stationary mode with vanishing values of the imaginary parts of the eigenvalues.
We find a new non-local multimode DCE represented by a multimode Bogoliubov transformation of the cavity mode and the photon band.
We show the practical advantage for the observation of DCE in that we can largely reduce the pump frequency when the cavity system is embedded in a narrow band photonic crystal with a bandgap.
\end{abstract}

\date{\today}

\maketitle

\section{Introduction}

A vacuum fluctuation is one of the most characteristic features of quantum mechanics, which has no classical analog \cite{MilonniBook}.
Lamb shift\cite{Lamb1947} and spontaneous emission \cite{Dirac1927,Weisskopf30ZeitPhys} are well-known examples of the effects of vacuum fluctuation on an atom.
The static Casimir effect is another example where an attractive force is working on electrically neutral bodies due to an exchange of a virtual photon surrounding a material\cite{Casimir48PR,Lamoreaux97PRL}.
In contrast to these effects, the dynamical Casimir effect (DCE) provides a more direct method to observe the quantum vacuum fluctuation. 
The rapid motion of the boundary of an electromagnetic field invokes  the conversion of localized virtual photons to real photons, which we can detect at a far distance\cite{Moore70JMP,Fulling1976,Dodonov2010,Nation12RMP}.
The DCE has also been intriguing because of its close relation to Hawking radiation and the Unruh effect\cite{Nation12RMP}.

Even with these interests, it has been difficult to experimentally observe the DCE because it requires a fast motion of the macroscopic body with almost the light velocity \cite{Moore70JMP,Fulling1976}.
Almost 40 years after the prediction by Moore, a few experiments to observe the DCE have succeeded with the use of a superconducting circuit to effectively change the boundary condition of the optical transmission line as well as the observation of the quantum nature of the emitted field, such as entangled photons and squeezing effect\cite{Wilson11Nature,Lahteenmaki2013a}.

Two exponential instabilities play a key role in the DCE.
The first is a quantum parametric amplification of a vacuum fluctuation, where a periodical change of an oscillator frequency enhances a virtual transition to yield a squeezed vacuum represented by the Bogoliubov transformation \cite{Carmichael07statisticalBook,Walls2008}.
As an external pump frequency comes close to twice the oscillator fundamental frequency, the bifurcation of the exponential amplification-deamplification  instability appears\cite{Mollow1986,Wu1987a,Kohler97PRE,Carmichael07statisticalBook,Walls2008}. 
The second is a resonance instability at which the irreversible transition from the amplified virtual photon to a real observable photon appears as a spontaneous photon emission to a free radiation field with exponential decay.
The point where the bifurcation of the resonance instability occurs is known as the {\it exceptional point}\cite{BenderPTsymmetry,MoiseyevBook}.
The stationary energy flow to the free radiation field emerges as a result of a balance between the parametric amplification and the dissipation processes with exponential growth and decay.

Since these two instabilities are dynamical processes, they are expected to be interpreted in a unified manner within a quantum dynamics.
Nevertheless, while the parametric instability has been well formulated by using the Bogoliubov transformation  \cite{Carmichael07statisticalBook,Walls2008}, the resonance instability has brought about a serious problem  as for how to derive the irreversibility based on the reversible microscopic principle of dynamics\cite{Prigogine1977,Petrosky91Physica,Prigogine1992,Prigogine1999528}.
Ordinary textbooks of quantum mechanics state that a hermitian Hamiltonian only possesses real eigenvalues, indicating no room for irreversibility.

Conventional descriptions of the dissipation processes in the DCE are input-output theories \cite{Collett84PRA,Ciuti06PRA} and a quantum master equation methods \cite{Carmichael87JOSAB,Kohler97PRE,Liberato09PRA}.
Since these theories are mostly based on the Markov approximation assuming an infinite bandwidth of the free radiation field, they are insufficient to describe the DCE photon emission to a narrow-bandwidth photonic crystal with a bandgap \cite{John1990,John1994,DeLiberato2014,Calajo2017,Rybin2017}.
Recent advances in hybrid quantum systems, such as optomechanical systems where a photon emission process is manipulated at a single photon level, require a theory of DCE taking into account a microscopic dissipation mechanism \cite{Xiang2013,Aspelmeyer2014,Settineri2019a}.

Recently, for the microscopic description of the dissipation process within a quantum mechanics, a new formalism, known as complex spectral analysis \cite{Petrosky00PRA,Karpov2000b,Petrosky01PRA,Ordoez2001} and non-Hemitian quantum mechanics \cite{Hatano97PRB,Bender98PRL,MoiseyevBook,BenderPTsymmetry}, have been independently developed.
In the complex spectral analysis, the functional space for a quantum state is extended to {\it  the rigged Hilbert space} where the dual functional space is equipped with a bi-complete and bi-orthonormal basis set\cite{Petrosky91Physica,Prigogine1992}, so that the time-evolution generator, Hamiltonian or Liouvillian, has  complex eigenvalues.
We have applied this theory to open quantum systems to study dissipation processes of a discrete quantum state interacting with a continuum with a finite bandwidth.
We have revealed that the decay is nonanalytically enhanced when the discrete state is located closely to the bandedge of the continuum, and, as a result, it shows a nonanalytic decay process\cite{Tanaka06PRB,Tanaka2007,Tanaka13PRA,Tanaka16PRA}.
Therefore, in order to describe DCE of a hybrid quantum system, it is important to take into account the effect of the energy-dependent self-energy. 

\begin{figure}
\begin{center}
\includegraphics[height=70mm,width=90mm]{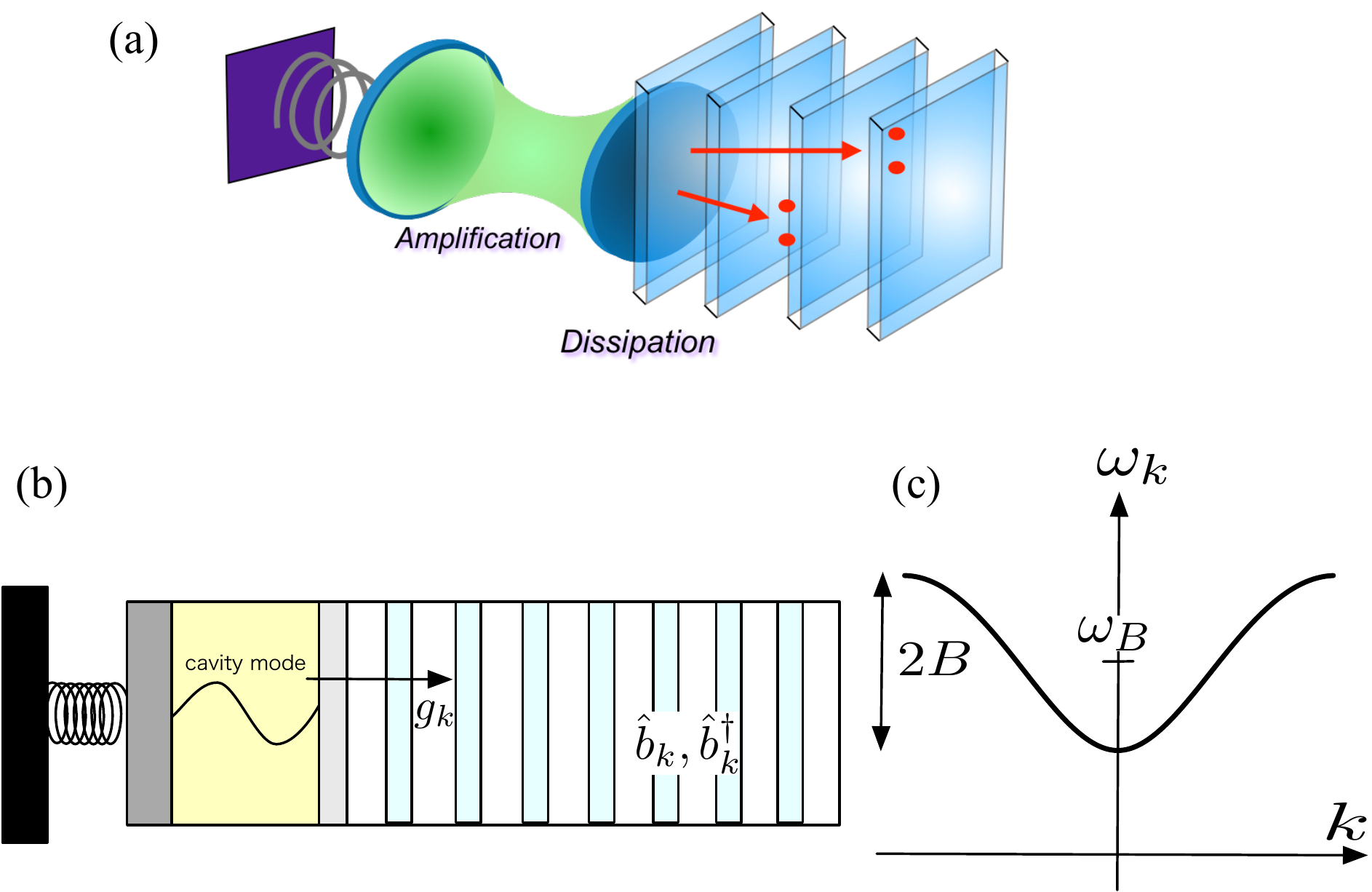}
\caption{(a) Optomechanical cavity interacting with a photonic crystal. (b) The frequency of a single cavity mode is periodically changed by mechanical pumping, and the cavity mode photon decays into one-dimensional photonic band. (c) Dispersion relation of one-dimensional photonic band with a bandwidth of $2B$ and the central frequncy $\omega_B$.}
\label{fig:hybrid}
\end{center}
\end{figure}

 In the present paper, we theoretically study the parametric amplification of a quantum vacuum of an optomechanical cavity interacting with a photonic band, where the mirror boundary is periodically moved by a classical external force, as shown in Fig.\ref{fig:hybrid}(a). 
The total system is composed of optomechanical cavity and photonic band states, and the time evolution of the operators of the canonical variables obeys the Heisenberg equation, where the generator of the time evolution is the Liouvillian.
With the use of Floquet method, we have transformed the time dependent problem to time-independent eigenvalue problem in Floquet space\cite{Sambe73PRA,Kohler97PRE,Grifoni1998}.
The non-Hermitian effective Liouvillian is derived in terms of the Brillouin-Wigner-Feshbach projection operator methods, where the microscopic dissipation process is rigorously taken into account with an energy-dependent self energy \cite{Feshbach62AnnalPhys,Rotter09JPhysA,Hatano2013,Kanki2017,Yamane18Symmetry}.
The complex eigenvalue problem of the effective Floquet-Liouvillian is solved to obtain new normal modes in terms of the multimode  Bogoliubov transformation, where the stationary mode is determined by the one with a vanishing imaginary part of its eigenvalue as a result of the balance between the parametric amplification and the dissipation.
We have found that we can stabilize the nonlocal mode, taking advantage of this balance when the cavity mode frequency lies in a photonic bandgap and that we can reduce the pump frequency to cause the DCE.

In Section \ref{Sec:Model}, we show the model of a hybrid quantum system consisting of optomechanical cavity and photonic crystal, and the total Hamiltonian for the system.
With the use of the Floquet method, we transform the Heisenberg equation to the time-independent complex eigenvalue problem of Floquet-Liouvillian.
The effective Floquet-Liouvillian is derived by using Brillouin-Wigner-Feshbach projection method, where the microscopic dissipation effect is rigorously taken into account in terms of the energy-dependent self-energy, associated with the detailed derivation in Appendix \ref{AppSec:FL}.
We show the calculated results in Section \ref{Sec:Results} where the competition between the parametric amplification and the dissipation may be clear in comparison with a phenomenological calculation.
The emergence of the nonlocal multimode DCE will be revealed as a result of the band edge effect of the photonic band.
We make a concluding remarks in Section \ref{Sec:Conclusion}.

%
%
%

\section{Model and Floquet-Liouvillian}\label{Sec:Model}

We consider  a hybrid quantum system consisting of optomechanical cavity and one-dimensional photonic crystal,  as shown in Fig.\ref{fig:hybrid}(a), where we assume a single cavity mode in the cavity and a one-dimensional photonic band represented by a tight-binding model for the photonic crystal as shown in Fig.\ref{fig:hybrid}(b).
The cavity mode decays into photonic band through the transmitting mirror on the one side of the cavity.
An external mechanical force oscillates a boundary mirror on the other side of the cavity with a frequency $\Omega$, which periodically changes the cavity mode frequency.

The total system is represented by the Hamiltonian\cite{Law94PRA,Law1995PRA} 
\begin{align}
\label{Hcav}
\hat H(t)=\omega_0 \hat a^\dagger \hat a+f(t)(\hat a+\hat a^\dagger )^2 + \int \omega_k  \hat b_k^\dagger \hat b_k dk + \int g_k(\hat a^\dagger \hat b_k+\hat b_k^\dagger \hat a) dk \;,
\end{align}
where $\hat a^\dagger$ ($\hat a$) and $\hat b_k^\dagger$ ($\hat b_k$) are the creation (annihilation) operators of the cavity mode and photonic band, respectively, and $\hbar=1$ is taken in the present paper. 
The effect of the external mechanical force on the cavity mode is represented by a change of  the cavity mode frequency in the second term  
\begin{align}
 f(t)=  f_0 \sin({\Omega t}+\theta)  \;,
 \end{align}
where $\Omega$, $f_0$, and $\theta$ are the pumping frequency, the amplitude, and the initial phase, respectively.
The photonic band is described by a one-dimensional tight-binding model whose dispersion relation is given by
\begin{align}
\omega_k=\omega_B-B\cos k \;.
\end{align}
as shown in Fig.\ref{fig:hybrid}(c).
The interaction of the cavity with the photonic band is described by the last term of (\ref{Hcav}), where the coupling strength for each $k$ mode is given by $g_k=g\sin k$ with a coupling constant $g$.
We adopt the rotating wave approximation as for the interaction between the cavity mode and the photonic band.

The Heisenberg's equation of the dynamical variables $\{ \hat a^\dagger  (t), \hat a (t), \{ b_k^\dagger (t),\hat b_k (t)\} \}$ are written by 
\begin{align}\label{Heisen1}
-i{d\over dt}
 \begin{pmatrix}
\hat a^\dagger \\
\hat a \\
\hat b_k^\dagger \\
\hat b_k\\
\vdots
\end{pmatrix} 
=
\begin{pmatrix}
\omega_0 +2 f(t) & 2 f(t)  & g_k & 0 & \hdots\\
-2 f(t) &-\omega_0-2f(t)  & 0 & -g_k& \hdots\\
g_k & 0 & \omega_k & 0 &\hdots\\
0 & -g_k & 0 &  -\omega_k  &\hdots\\
\vdots& \vdots&\vdots&\vdots&\ddots\\
\end{pmatrix}
 \begin{pmatrix}
\hat a^\dagger \\
\hat a \\
\hat b_k^\dagger \\
\hat b_k\\
\vdots
\end{pmatrix}\equiv {\cal L}(t)  
 \begin{pmatrix}
\hat a^\dagger \\
\hat a \\
\hat b_k^\dagger \\
\hat b_k\\
\vdots
\end{pmatrix}  \;.
\end{align}
The generator of time evolution is obtained from the commutation relation with the Hamiltonian which is now represented by a matrix in (\ref{Heisen1})  which we shall call Liouvillian matrix ${\cal L}(t)$.
Note that the virtual transition interactions appear in the off-diagonal matrix elements hybridizing the $\hat a^\dagger$ and $\hat a$ of the cavity modes.
As a result, these modes at time $t$  are represented  by the Bogoliubov transform which changes a bare vacuum state to a squeezed vacuum state\cite{LoudonBook,Walls2008}.
In (\ref{Heisen1}),  the column vector of the operators represents the field operator of the radiation field
\begin{align}
|\hat \Psi(t)\>\equiv \hat a^\dagger(t) |\varphi_{a^*}\> +\hat a(t) |\varphi_a \>+\int \left(\hat b_k^\dagger(t)  |\varphi_{k^*}\>+\hat b_k (t) |\varphi_{k}\> \right)dk \;,
\end{align}
where $|\phi_j\>$'s are time-independent basis used in the second quantization of the radiation field\cite{Cohen1989photonBook}.
We assume these mode basis form an orthonormal complete basis set:
\begin{align}
\<\phi_i|\phi_j\>=\delta_{i,j} \;,\; 1=|\varphi_a\>\<\varphi_a|+|\varphi_{a^*}\>\<\varphi_{a^*}|+\int dk \left( |\varphi_k\>\<\varphi_k|+|\varphi_{k^*}\>\<\varphi_{k^*}| \right)\;.
\end{align}

The field operator can be also expanded by the solutions of  scalar wave equations 
\begin{align}\label{WaveEq}
-i {d  \over dt}|\Psi_\xi (t)\>= {\cal L}(t)|\Psi_\xi (t)\> \;,
\end{align}
where ${\cal L}(t)$ is the Liouvillian matrix in (\ref{Heisen1}).
Then the field operator is expanded by the mode functions $|\Psi_\xi (t)\>$  as in Ref.\cite{Mollow1986}:
\begin{align}
|\hat\Psi (t)\>=\sum_\xi \hat\Psi_\xi |\Psi_\xi(t)\> \;.
\end{align}

Since ${\cal L}(t)$ is time-periodic, the wave equation (\ref{WaveEq}) can be solved by the Floquet method.
We can write the solution as \cite{Sambe73PRA,Kohler97PRE,Grifoni1998}
\begin{align}\label{Psixit}
|\Psi_\xi(t)\>=e^{i z_\xi t}|\Phi_\xi(t)\> \;,\; |\Phi_\xi(t+T)\>=|\Phi_\xi(t)\> \;,
\end{align}
where $T=2\pi/\Omega$,  $z_\xi$ is the {\it quasi}-eigenvalue of the Floquet-Liouvillian, and the $|\Phi_\xi(t)\>$ is the corresponding eigenstate satisfying
\begin{align}\label{EVLt}
{\cal L}_{\rm F}(t)|\Phi_\xi(t)\>\equiv \left[ {\cal L}(t)-i{d\over dt}\right] |\Phi_\xi(t)\>=z_\xi|\Phi_\xi(t)\> \;,\; (z_\xi  \text{ mod } \Omega).
\end{align}

In terms of the Floquet transform,  the continuous time variable $t$ is transformed to the discrete Floquet mode variables, by means of which  the time-dependent differential equation (\ref{WaveEq}) turns to time-independent complex eigenvalue problem of the Floquet-Liouvillian as shown in Appendix \ref{AppSec:FL}.
It is found that the virtual transition interactions couples the creation and annihilation modes of the cavity belonging to a next neighbor Floquet modes,  making ${\cal L}_{\rm F}$ non-Hermitian.

Under the condition of 
\begin{align}
\Omega-2B \gg f_0 \;,
\end{align}
which is satisfied for the narrow photonic band and small external amplitude, we can restrict ourselves to $\{ |a^*,1)\!\>, |a,0)\!\> \}$ space.
By using the projection operator methods\cite{Feshbach62AnnalPhys}, we derive the effective Floquet Liouvillian in terms of the cavity modes of $\{ |a^*,1)\!\>, |a,0)\!\> \}$ given by
\begin{align}\label{Leff}
{\cal L}_{\rm eff}(z)=\begin{pmatrix}
\omega_0-\Omega+g^2 \sigma(z+\Omega)  & -if_0 e^{i\theta} \\
-if_0e^{-i\theta} &  - \omega_0 -g^2 \sigma(-z)
\end{pmatrix}  \;,
\end{align}
as shown in Appendix \ref{AppSec:FL}.
For the present one-dimensional photonic band, the self-energy is analytically obtained by
\begin{align}
\sigma(z)\equiv \int_{-\pi}^\pi {\sin^2k\over z-\omega_k} dk =z-\sqrt{z^2-B^2} \;.
\end{align}

As seen from  (\ref{Leff}), the two competing instabilities mentioned in the Introduction clearly appear in the effective Liouvillian.
The virtual transition in the off-diagonal elements and the complex self-energy in the diagonal elements represent the parametric instability and the resonance instability, respectively.
Therefore, the effective Liouvillian describes the exponential instabilities in the DCE from a unified point of view.

\begin{figure}
\begin{center}
\includegraphics[height=50mm,width=110mm]{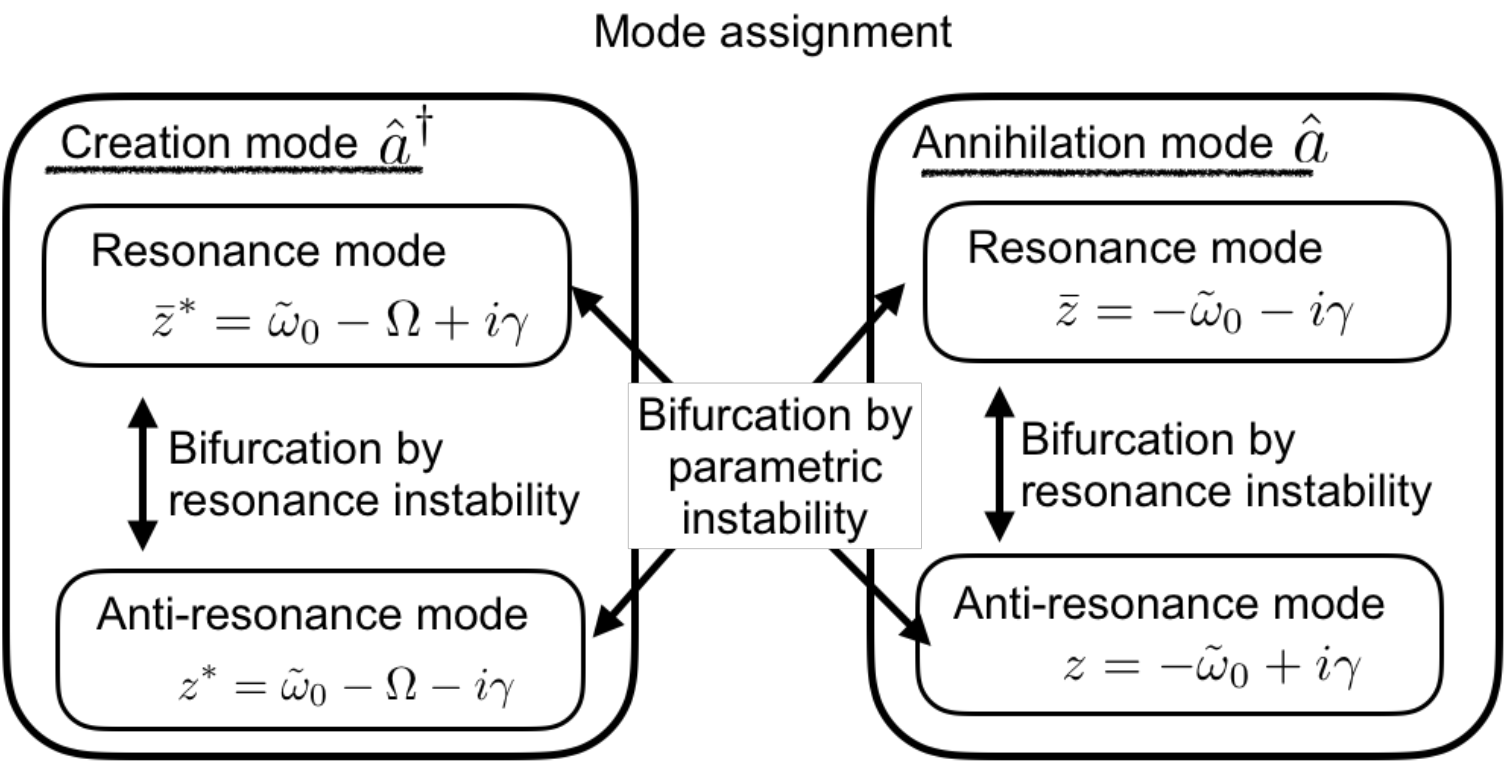}
\caption{Characteristic origin of the eigenmodes of the system. }
\label{fig:Modes}
\end{center}
\end{figure}

Now the complex eigenvalue problem of the total Floquet-Liouvillian is reduced to the one of the effective Liouvillian, which reads
\begin{align} \label{LeffEV}
{\cal L}_{\rm eff}(z_\xi)|\varphi_\xi)\!\>=z_\xi |\varphi_\xi )\!\> \;,\; \<\!( \tilde\varphi_\xi|{\cal L}_{\rm eff}(z_\xi)=z_\xi \<\!( \tilde\varphi_\xi|\;.
\end{align}
The complex eigenstates form a bi-orthonormal bi-complete basis set \cite{Petrosky91Physica}:
\begin{align}
1=\sum_\xi|\varphi_\xi)\!\>\<\!(\tilde\varphi_\xi|\;,\; \<\!(\tilde\varphi_\xi|\varphi_{\xi'})\!\>=\delta_{\xi,\xi'} \;.
\end{align}
The complex eigenvalues $z_\xi$ in (\ref{LeffEV}) are obtained by solving the dispersion equation 
\begin{align}\label{dispersionEq}
\left(z+\omega_0+g^2\sigma(-z)\right)\left(z-\omega_0+\Omega-g^2\sigma(z+\Omega)\right) +f_0^2=0 \;.
\end{align}
By squaring the square root functions, this dispersion equation turns to a eighth-order polynomial equation which has been numerically solved.
Among the solutions, we found the four solutions continuously connecting to the unperturbed modes.
The physical origin of the four solutions are assigned to a mixture of the resonance and antiresonance modes for each of cavity creation and annihilation modes, as shown in Fig.\ref{fig:Modes}.

In terms of the complex spectral analysis, we can define the stationary mode based on the microscopic dynamics as the mode whose eigenvalue has a vanishing value of the imaginary part, i.e., ${\rm Im}z_j=0$, without relying on Markov approximation.
In the next section, we shall see the profound effect of the band edge of photonic band on the generation of the real photon emission in the dynamical Casimir effect of the hyrbrid quantum system.

\section{Results}\label{Sec:Results}

\begin{figure}
\begin{center}
\includegraphics[height=80mm,width=130mm]{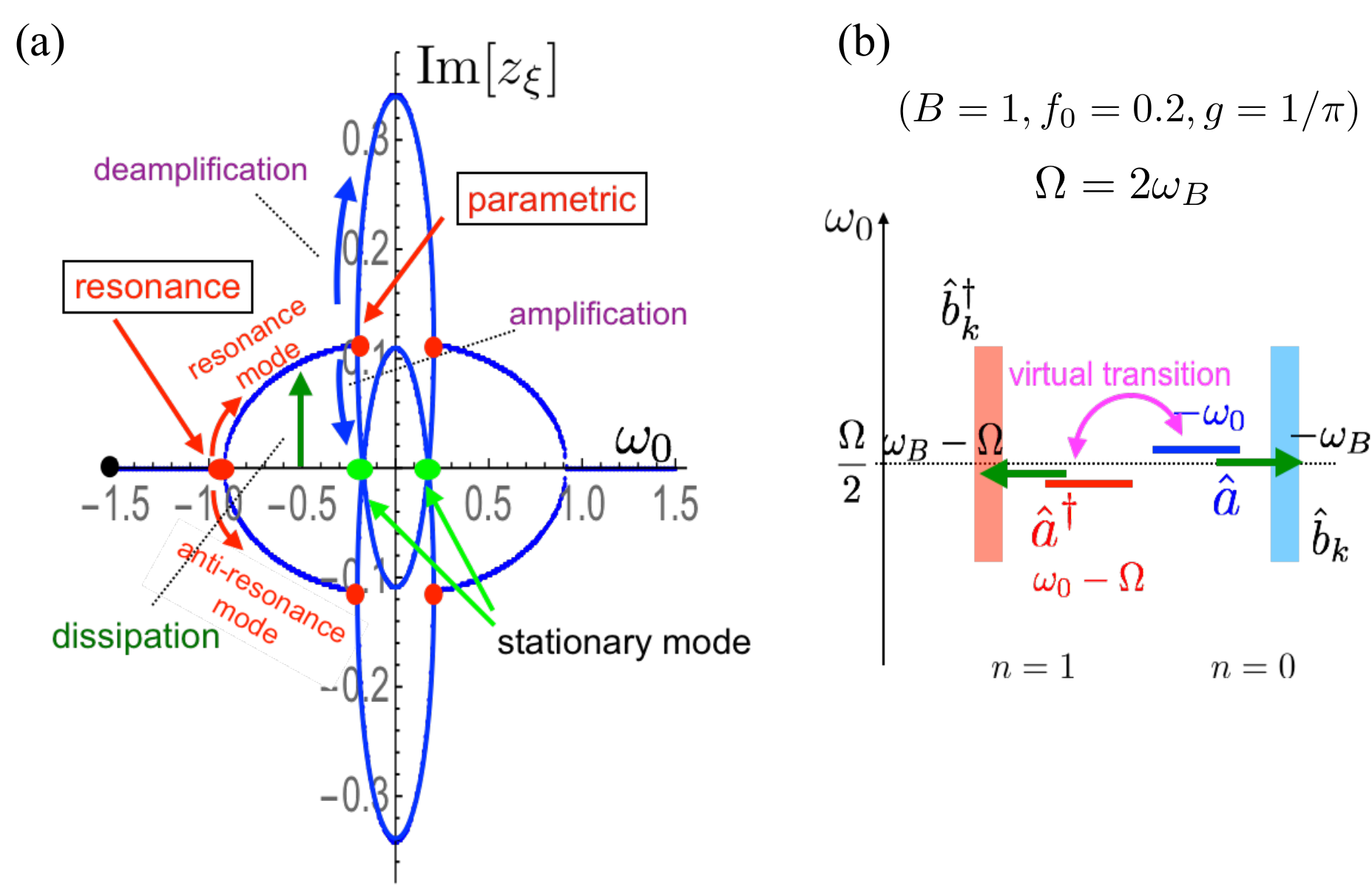}
\caption{Calculated results for $\Omega=2\omega_B$, where $B=1$, $f_0=0.2$, and $g=1/\pi$ are fixed: (a) Imaginary part of the complex eigenvalues of ${\cal L}_{\rm eff}$ as a function of $\omega_0$, where the origin of the horizontal axis taken at $\omega_B$. The bifurcation points are indicated by the red filled circles, and the stationary point is indicated by the green filled circle.(b) A schematic picture of the Floquet-Liouvillian around the stationary point.}
\label{fig:Res1}
\end{center}
\end{figure}

In order to show the effect of the balance between the parametric amplification and the dissipation, we show the imaginary parts of the eigenvalues of ${\cal L}_{\rm eff}$ for $\Omega=2\omega_B$ in Fig.\ref{fig:Res1} where we change the cavity frequency $\omega_0$ while the values of $f_0=0.2$ and   $g=1/\pi$ are fixed.
As shown in Fig.\ref{fig:Res1}(b), the neighboring Floquet-photonic bands are overlapped.

As $\omega_0$ increases, we encounter the bifurcation of the resonance instability at $\omega_0\simeq\omega_B-B\simeq -0.9$, where the cavity mode becomes resonant with the photonic band, resulting in the bifurcation to resonance and anti-resonance modes as in the other decaying systems\cite{Tanaka06PRB,Tanaka16PRA,Fukuta17PRA,Tanaka18META}.
In this figure, a positive vertical direction indicates a state decaying since $|\Psi_\xi(t)\>\propto \exp[-({\rm Im}z_\xi )t]$ as shown in (\ref{Psixit}).
With a further increase of $\omega_0$, the frequencies of the creation and annihilation cavity modes come close and the effect of the virtual transition between them becomes significant.
Then we encounter the second bifurcation of parametric instability at $\omega_0\simeq \omega_B-f_0$, where the downward and upward branches corresponds to the parametric amplification and deamplification, respectively.
As we further increase $\omega_0$, we reach to the stationary point where ${\rm Im}z_\xi=0$,  as a result of a balance between the parametric amplification and the dissipation effects, as indicated by the green filled circle.
At this point, the stationary energy flow through the cavity from the external pump to the photonic band is achieved with the spontaneous photon pair emission.
The figure clearly demonstrates that this stationary DCE has been determined by taking into account the microscopic dissipation process.

\begin{figure}
\begin{center}
\includegraphics[height=55mm,width=130mm]{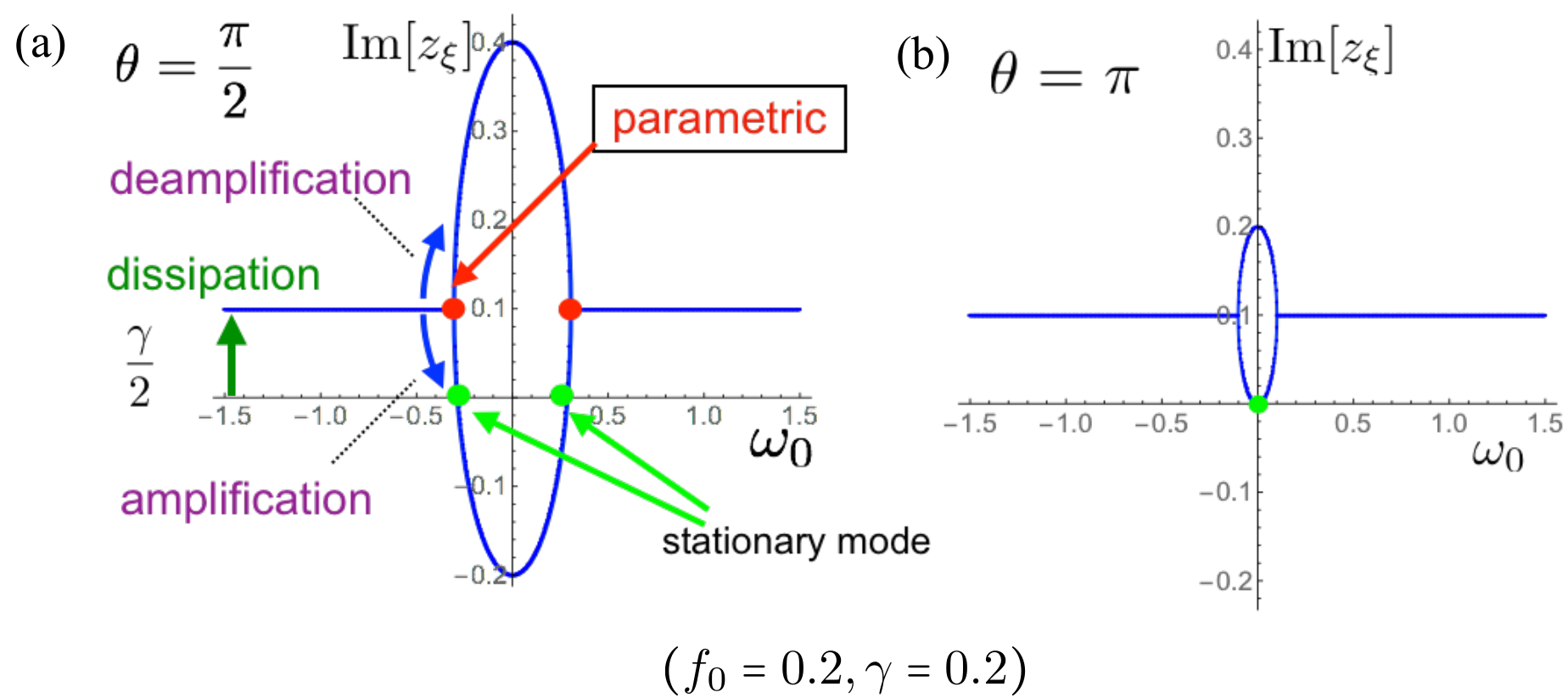}
\caption{Imaginary part of the complex eigenvalues of the phenomenological model  as a function of $\omega_0$ for the values of $f_0=0.2$ and $\gamma=0.2$, where the origin of the horizontal axis at $\omega_0=\Omega/2$. (a) $\theta=\pi/2$,  and (b) $\theta=\pi$.}
\label{fig:ResPh}
\end{center}
\end{figure}

Here, we compare the present results with a phenomenological theory.
We write a phenomenological effective Floquet-Liouvillian
\begin{align}
{\cal L}_{\rm ph}(z)=\begin{pmatrix}
\omega_0-\Omega+i{\gamma\over 2}  & -if_0 e^{i\theta}-i{\gamma\over 2}  \\
-if_0 e^{-i\theta}-i{\gamma\over 2}  &  - \omega_0 +i{\gamma\over 2} 
\end{pmatrix} 
\end{align}
 consistent with an equation of motion of a forced damped oscillator
\begin{align} 
 \ddot{x}+\gamma \dot{x}+\omega(t)^2 x=0\;,
 \end{align}
where 
\begin{align}
 \omega^2 (t) =\omega_0^2\left( 1+ {2f_0\over \omega_0}\sin ( \Omega t +\theta) \right) \;,
 \end{align}
and $\gamma$ is a phenomenological dissipation constant.
The complex eigenvalue is immediately obtained by
\begin{align}\label{Phenomzj}
z_\xi=-{\Omega\over 2}+ i {\gamma \over 2}  \pm \sqrt{ \left( \omega_0-{\Omega\over 2}  \right)^2- \left( f_0^2 +\left({\gamma\over 2}\right)^2+\gamma f_0 \cos \theta\right) }\;.
\end{align}
In Fig.\ref{fig:ResPh},  we show the imaginary part of the above solutions.
Since this phenomenological model assumes  a flat-band radiation with infinite bandwidth, the resonance bifurcation does not appear, though we see  the stationary mode as a balance between the constant dissipation and parametric amplification.
What seems unphysical in (\ref{Phenomzj}), however,  is its initial phase dependence of the results as shown in Fig.\ref{fig:ResPh}(a) and (b).
It is irrational that the stationary point depend on the initial phase of the external field. 
Contrary to this unphysical results, the correct result based on the microscopic model does not depend on the phase factor, as seen from (\ref{dispersionEq}).

\begin{figure}
\begin{center}
\includegraphics[height=110mm,width=130mm]{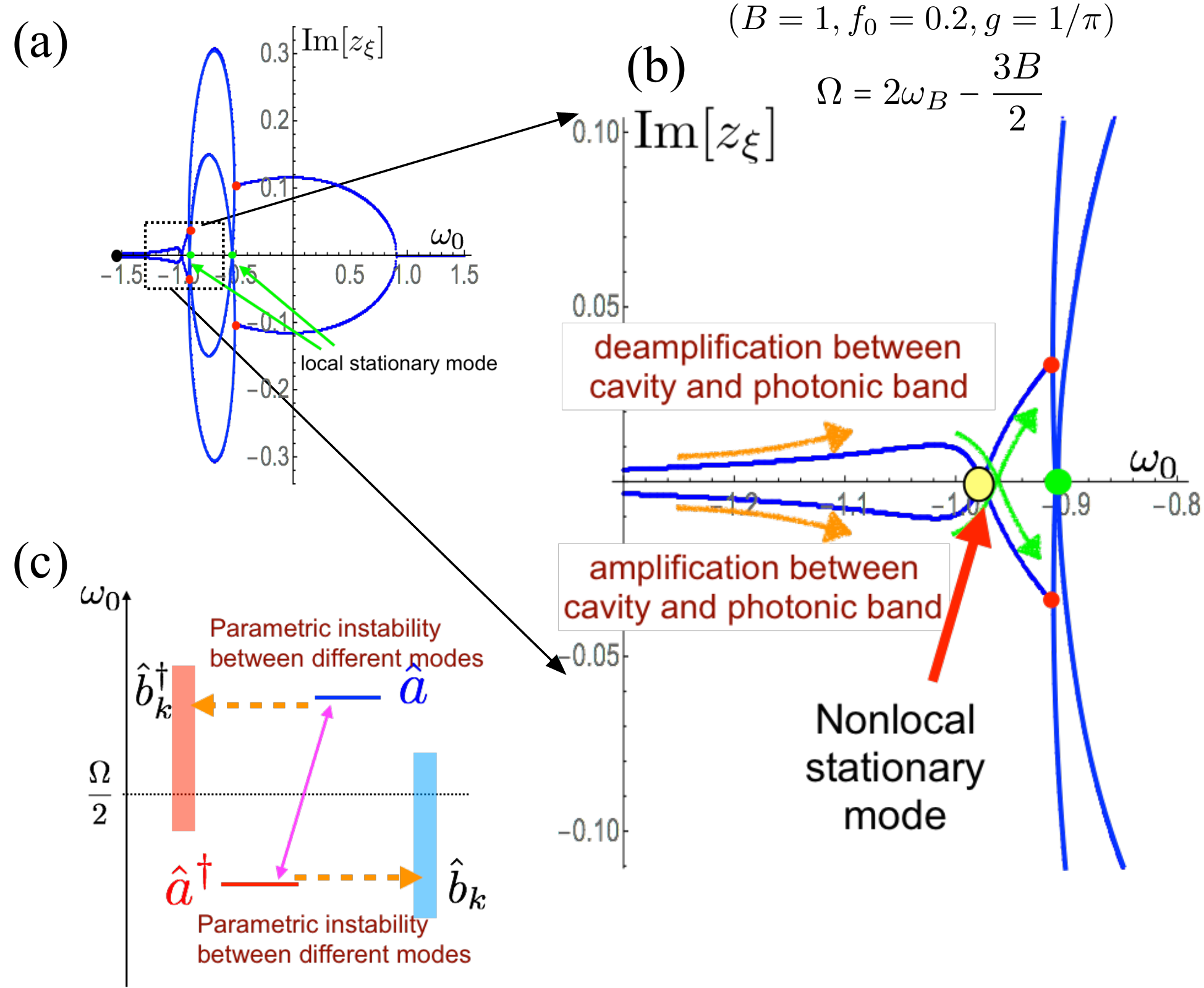}
\caption{Calculated results for $\Omega=2\omega_B-\Delta$, where $\Delta=3B/2$, $B=1$, $f_0=0.2$, and $g=1/\pi$ are fixed: (a) Overall picture of the imaginary part of the complex eigenvalues of ${\cal L}_{\rm eff}$ as a function of $\omega_0$, where the origin of the horizontal axis taken at $\Omega/2$. The bifurcation points are indicated by the red filled circles, and the stationary point is indicated by the green filled circle. (b) Expanded picture of (a) around the band edge. Non-local multimode stationary point is indicated by the yellow filled circle. (c) A schematic picture of the Floquet-Liouvillian around the stationary point.}
\label{fig:Res2}
\end{center}
\end{figure}

The band edge effect is more pronounced when lowering the external pumping frequency.
In Fig.\ref{fig:Res2}, we show the results  for $\Omega=2\omega_B-\Delta$ with $\Delta=3B/2$ so that the neighboring Floquet photonic bands are shifted by $\Delta$, as shown in Fig.\ref{fig:Res2}(c), where other parameters are fixed at the same values of Fig.\ref{fig:Res1}.
The overall behavior of ${\rm Im}z_\xi$ is shown in Fig.\ref{fig:Res2}(a), where we have seen essentially the same stationary points, indicated by the green filled circles, as a result of the balance between the resonance instability and the parametric amplification of the cavity modes. 

But in the region where the cavity mode frequency is located around the band edge, i.e. $\omega_0\simeq \omega_B-B$, we can see the other type of the stationary point.
We show the result in the expanded scale in Fig.\ref{fig:Res2}(b), and the schematic picture of the Floquet-Liouvillian in this region is shown in Fig.\ref{fig:Res2}(c).
In this region, the creation (annihilation) cavity mode is in resonance with the annihilation (creation) mode of the photonic band.
Even though there is no direct coupling between them, there is an indirect coupling via the virtual transition of the cavity modes, as shown in Fig.\ref{fig:Res2}(c).
Consequently, the eigenmode of the total system is represented by the {\it multimode-Bogoliubov transformation} of the cavity mode and the {\it continuous} photonic band \cite{LoudonBook}.
This mixing of the creation and annihilation modes between them induces the {\it non-local multimode parametric instability}, as shown by the orange arrows in Fig.\ref{fig:Res2}(b).

As $\omega_0$ further increases so that the creation (annihilation) cavity mode is in resonance with the corresponding photonic band modes, the multimode parametric instability is suppressed by the dissipation effect.
As a result of the balance between them, the nonlocal stationary point emerges, which is indicated by the yellow filled circle in Fig.\ref{fig:Res2}(b). 
This stationary mode is characteristically different from the previous one, because here the nonlocal entanglement between the cavity photon and photonic band photons appears.

\section{Concluding remarks}\label{Sec:Conclusion}

In this paper, we have studied the parametric amplification of a quantum vacuum in the optomechanical cavity interacting with a one-dimensional photonic crystal.
The effective non-Hermitian Floquet-Liouvillian has been derived from a Heisenberg equation of the total system by  using the Floquet method and the Brillouin-Wigner-Feshbach projection method, where we have taken into account a microscopic dissipation mechanism in terms of the energy-dependent self energy. 
The non-hermitian effective Floquet-Liouvillian reveals the competing instabilities of parametric amplification and dissipation due to the virtual transition and the resonance singularity, respectively.
The emergence of the stationary mode has been identified as a result of the balance between the two instabilities, where the eigenmode of the Liouvillian is represented by the Bogoliubov transformation.
The photonic band edge effect is prominent when the cavity mode frequency is close to the band edge.
In this case, the indirect coupling between the cavity mode and photonic band via the virtual transition of the cavity modes yields the nonlocal stationary mode which is represented by the multimode-Bogoliubov tranformation of  the cavity mode and photonic band.

\begin{figure}
\begin{center}
\includegraphics[height=50mm,width=80mm]{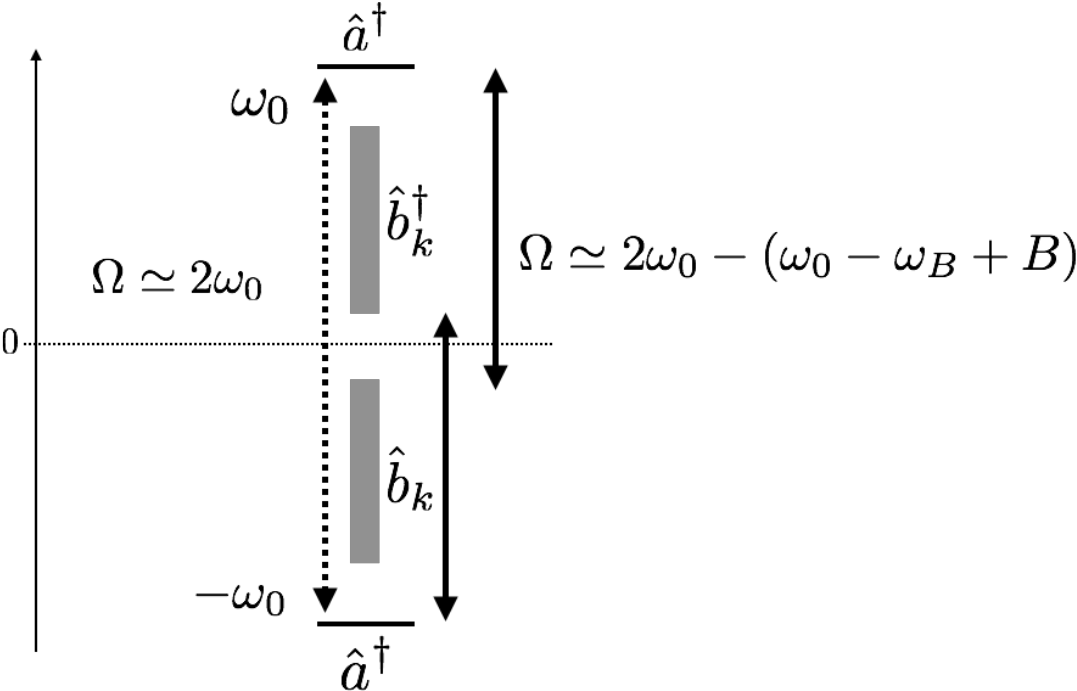}
\caption{Calculated results for $\Omega=2\omega_B$: (a) Imaginary part of the complex eigenvalues of ${\cal L}_{\rm eff}$ as a function of $\omega_0$. (b) Floquet-Liouvillian level scheme.}
\label{fig:Reduction}
\end{center}
\end{figure}

Lastly, we would emphasize a practical advantage of the present model for the observation of the DCE.
A major obstacle for the observation of the DCE is the difficulty to move the boundary with almost twice of the cavity frequency, $\Omega\simeq 2\omega_0$.
But as seen from the preceding section, the pump frequency can be lowered to cause the multimode parametric amplification, where the creation (annihiation) cavity mode are mixed with the annihilation (creation) mode of the photonic band.
We have found the conditions for the appearance of the multimode stationary state.
First,  
\begin{align}
\omega_0>\omega_B+B 
\end{align}
is necessary for that the cavity mode is stable in spite of the pumping field.
Then, under this condition the pumping field with 
\begin{align}
\Omega\sim 2\omega_0-(\omega_0-\omega_B+B) < 2\omega_0-2B \;, \; 
\end{align}
far smaller than $2\omega_0$, enables the indirect virtual coupling between the cavity mode and photonic band as shown in Fig.\ref{fig:Reduction}.
Therefore, the major obstacle can be diminished.

In the present work, we have found three different types of the stationary modes.
One is a completely stationary state well below the two bifurcation thresholds, where a cavity squeezed vacuum state is associated with a localized virtual photon cloud of the photonic band, and the periodic Rabi oscillation happens between the cavity squeezed vacuum and the virtual photon cloud.
The second one is the multimode DCE, where the stationary spontaneous photon emission to the photonic band happens with the two-photon entanglement between the cavity mode and the photonic band.
The third one is the ordinary DCE, where the entangled cavity photon pair is emitted to the photonic band.
We can observe the two-photon entanglement of the emitted photons by a quantum correlation observation, such as the homodyne detection method\cite{LoudonBook,Walls2008}.
We will show the study of the real-time dynamics of these photon emission process in the forthcoming paper.




\acknowledgments

We are very grateful T. Petrosky, R. Passante, H. Yamane, Y. Kayanuma, K. Mizoguchi, K. Noba, S. Garmon, and M. Domina for fruitful discussions.

%
%

%

\appendix
\section{Floquet-Liouvillian complex eigenvalue problem and effective operator}\label{AppSec:FL}

In this section, we briefly review the Floquet method according to Ref.\cite{Shirley1965,Sambe73PRA,Grifoni1998}, and derive the effective Floquet-Liouvillian given in (\ref{Leff}).
The composite space ${\cal F}\equiv {\cal R}\otimes{\cal T}$ is composed of the configuration space  ${\cal R}$  and the space  ${\cal T}$  of periodic functions in time with period $T$ \cite{Grifoni1998}.
In ${\cal T}$-space, any periodic function of $t$  is represented as a vector
$f(t)\equiv (t|f)$, 
where the time basis is an eigenstate of a time operator
$\hat t|t)=t|t) $,
and the conjugate operator is given by
$\hat p_t\equiv i {\partial/\partial t}$ \;.
The eigenstate of $\hat p_t$ is given by
\begin{align}\label{AppEq:kappan}
 |\kappa_n)={1\over T}\int_0^T e^{i\kappa_n t}|t) dt  \;,
\end{align}
where $\kappa_n=n\Omega=2\pi n/T\;, (n=0,\pm 1,\pm2, \cdots)$,
satisfying
\begin{align}
\hat p_t|\kappa_n)=\kappa_n|\kappa_n) \;.
\end{align}
The time basis is given by the transformation of
\begin{align}
|t)=\sum_{n=-\infty}^\infty e^{-i\kappa_n t}|\kappa_n) \;.
\end{align}
Together with the basis of ${\cal R}$-space, the complete orthonormal basis set in the ${\cal F}$-space is formed by $\{ |\varphi_j,t)\!\>\}$  or $\{ |\varphi_j,\kappa_n)\!\>\}$ $(j=a,a^*,k,k^*)$ in terms of the time- or Floquet-mode-representations, respectively, where $|\cdot,\cdot )\!\>$ denotes a vector in the $\cal F$-space.
In this paper, we have abbreviated as  $j\equiv \varphi_j$ and $n\equiv\kappa_n$.
These basis satisfy the complete-orthonormality
\begin{align}
1={1\over T} \sum_j \int_0^T dt |j,t)\!\>\<\!(j,t| \;, \; \<\!(j,t|j',t')\!\>=T\delta(t-t')\delta_{j,j'} \;,
\end{align}
in terms of $\{ |j,t)\!\>\}$, or  
\begin{align}
1=\sum_{n=-\infty}^\infty \sum_j  |j,n)\!\>\<\!(j,n| \;, \; \<\!(j,n|j',n')\!\>=\delta_{n,n'}\delta_{j,j'} \;,
\end{align}
 in terms of $\{ |j,n)\!\>\}$ basis set.

Using the transform of (\ref{AppEq:kappan}),  the complex eigenvalue problem of the Floquet-Liouvillian 
\begin{align}\label{AppEq:LF}
{\cal L}_{\rm F}|\Phi_\xi)\!\>=z_\xi|\Phi_\xi)\!\>
\end{align}
 is transformed to time-independent eigenvalue problem in terms of the Floquet-mode representation.
 The matrix structure of ${\cal L}_{\rm F}$  is given by
\begin{align}
{\cal L}_{\rm F}=\begin{array}{c|c c|c c| c c|c c|}
\hline
& \multicolumn{4}{|c|}{ n=1}   &  \multicolumn{4}{|c|}{ n=0}    \\
 &  a^* & a  & b_k^* & b_k  & a^*  & a  & b_k^* & b_k  \\ \hline
a^* &  \omega_0-\Omega & 0 & g_k & 0 & -if_0 e^{i\theta} & -if_0 e^{i\theta}  & 0 & 0 \\
a &  0 & -\omega_0-\Omega & 0 &-g_k  &   if_0 e^{i\theta} &    if_0 e^{i\theta}& 0  & 0 \\ \hline
b^* & g_k & 0 &\omega_k-\Omega  & 0 &0  &0  &0  &0  \\
b_k & 0 &  -g_k & 0 &-\omega_k-\Omega  &0  & 0 & 0 & 0 \\ \hline
 a^* &  if_0 e^{-i\theta}  & if_0 e^{-i\theta} &0  & 0 & \omega_0  &0  & g_k &0  \\
 a &- if_0 e^{-i\theta}   &- if_0 e^{-i\theta} & 0&  0&  0&   -\omega_0 &0 &-g _k \\ \hline
b_k^* & 0 &0  & 0 & 0 &g_k  &0  &\omega_k  &0  \\
b_k & 0 &0  &0  &0  &0  &-g_k  &0  &-\omega_k  \\ \hline 
\end{array} \;,
\end{align}
where we show the matrix only for the $n=0$ and $n=1$ Floquet modes and a particular $k$ mode of the photonic band, for simplicity.
It is found that the virtual transition interactions couples the creation and annihilation modes of the cavity belonging to a next neighbor Floquet modes:
\begin{align}
&\<\!(a^*,1|{\cal L}_{\rm F}|a,0 )\!\>=-i f_0e^{i\theta} \;,\; \<\!(a,1|{\cal L}_{\rm F}|a^*,0 )\!\>=i f_0e^{i\theta} \;,\;  \notag\\
&\<\!(a^*,0|{\cal L}_{\rm F}|a,1 )\!\>=i f_0e^{-i\theta} \;,\; \<\!(a,0|{\cal L}_{\rm F}|a^*,1 )\!\>=-i f_0e^{-i\theta}  \;,
\end{align}
which make ${\cal L}_{\rm F}$ non-Hermitian.

Under the condition of 
\begin{align}
\Omega-2B \gg f_0 \;,
\end{align}
which is satisfied for the narrow photonic band and small external amplitude, we can restrict ourselves to $\{ |a^*,1)\!\>, |a,0)\!\> \}$ space.
Using the projection operators \cite{Yamane18Symmetry} of 
\begin{align}
{\cal P}\equiv   |a^*,1)\!\>\<\!(a^*,1| +|a,0)\!\>\<\!(a,0|  \;,\; {\cal Q}\equiv 1-{\cal P}\;,
\end{align}
we have reduced the complex eigenvalue problem of the total Floquet-Liouvillian (\ref{AppEq:LF}) to an eigenvalue problem of an effective Floquet Liouvillian in terms of the cavity modes of $\{ |a^*,1)\!\>, |a,0)\!\> \}$  as 
\begin{align}\label{AppEq:LeffEV}
{\cal L}_{\rm eff}(z_\xi){\cal P}|\Phi_\xi)\!\>=z_\xi{\cal P}|\Phi_\xi)\!\> \;,
\end{align}
where the effective Floquet-Liouvillian is defined by\cite{Petrosky91Physica,Tanaka16PRA,Fukuta17PRA,Yamane18Symmetry,Tanaka18META}
\begin{align}
{\cal L}_{\rm eff}(z)\equiv {\cal P}{\cal L}_{\rm F} {\cal P}+ {\cal P}{\cal L}_{\rm F} {\cal Q}{1\over z-{\cal Q}{\cal L}_{\rm F}{\cal Q}}  {\cal Q}{\cal L}_{\rm F} {\cal P} \;.
\end{align}
The effective Floquet-Liouvillian is represented by a two-by-two matrix in terms of  $\{ |a^*,1)\!\>, |a,0)\!\> \}$
\begin{align}
{\cal L}_{\rm eff}(z)=
\begin{array}{c|c c|}
 &  |a^*,1)\!\> & |a,0)\!\>   \\ \hline
\<\!(a^*,1| &  \omega_0-\Omega +g^2\sigma(z+\Omega) &  -if_0 e^{i\theta}   \\
\<\!(a,0| &  - if_0 e^{-i\theta}  &  -\omega_0 -g^2\sigma(-z)\\ \hline 
\end{array} \;,
\end{align}
where the self-energy is analytically obtained for the interaction of the cavity mode with the one-dimensional photonic crystal as
\begin{align}
\sigma(z)\equiv \int_{-\pi}^\pi {\sin^2k\over z-\omega_k} dk =z-\sqrt{z^2-B^2} \;.
\end{align}

It should be noted that the effective Floquet-Liouvillian depends on its eigenvalue so that the eigenvalue problem should be solved in a self-consistent manner so as to correctly describe the microscopic dissipation mechanism.
The eigenvalues have been obtained by solving the dispersion equation (\ref{dispersionEq}) which has been derived from (\ref{AppEq:LeffEV}).








\end{document}